\documentclass{pasj01}
\draft 
\Received{$\langle$reception date$\rangle$}
\Accepted{$\langle$acception date$\rangle$}
\Published{$\langle$publication date$\rangle$}
\begin{document}

\title{Neutral iron line in the supernova remnant IC\,443 and implications for MeV cosmic rays}
\author{Kumiko K. \textsc{Nobukawa}\altaffilmark{1}, Arisa \textsc{Hirayama}\altaffilmark{1}, Aika \textsc{Shimaguchi}\altaffilmark{1}, \\ Yutaka \textsc{Fujita}\altaffilmark{2}, Masayoshi \textsc{Nobukawa}\altaffilmark{3}, and Shigeo \textsc{Yamauchi}\altaffilmark{1} }%
\altaffiltext{1}{Department of Physics, Faculty of Science, Nara Women's University, Kitauoyanishi-machi, Nara, Nara 630-8506, Japan}
\altaffiltext{2}{Department of Earth and Space Science, Graduate School of Science, Osaka University, Toyonaka, Osaka 560-0043, Japan}
\altaffiltext{3}{Faculty of Education, Nara University of Education, Takabatake-cho, Nara, Nara 630-8528, Japan}
\email{kumiko@cc.nara-wu.ac.jp}


\maketitle

\begin{abstract}
We report a discovery of bright blob-like enhancements of an Fe~\emissiontype{I} K$\alpha$ line in the northwest and the middle of the supernova remnant (SNR) IC\,443. The distribution of the line emission is associated with molecular clouds interacting with the shock front, and is totally different from that of the plasma. The Fe~\emissiontype{I} K$\alpha$ line has a large equivalent width.  The most plausible scenario for the origin of the line emission is that the MeV protons accelerated in the shell leak into the molecular clouds and ionize the Fe atoms therein. 
The observed Fe~\emissiontype{I} K$\alpha$ line intensity is consistent with the prediction of a theoretical model, in which MeV protons are accelerated along with GeV and TeV protons at the SNR.
\end{abstract}

\section{Introduction}
Supernova remnants (SNRs) are believed to be an acceleration site of Galactic cosmic rays. GeV and very-high energy (VHE) gamma-rays have been detected from many SNRs (e.g. \cite{Abdo09, HESS18}) and have contributed to promote our understanding of high-energy cosmic rays (in the GeV--TeV band; hereafter HECRs)  in SNRs. 
 According to the established acceleration theory, suprathermal particles are injected into the acceleration mechanism and are  accelerated to be low-energy cosmic rays (hereafter LECRs) and finally HECRs. In this context, LECRs have been a missing link of the particle acceleration in SNRs; there have been very limited information on LECRs so far in contrast to HECRs. LECRs are easily affected by solar modulation, and thus it is hard to  observe them inside the solar system. A direct and robust observation is conducted only by the Voyager mission \citep{Cummings16}. The ionization rate has provided  the unique indirect information on LECRs (e.g. \cite{Indriolo12}).    

Recently, X-ray observations have demonstrated that an Fe~\emissiontype{I}~K$\alpha$ line can be a powerful probe to investigate LECRs.  Neutral Fe atoms in cold material are ionized by LECRs to radiate the Fe~\emissiontype{I} K$\alpha$ line at 6.4~keV.  Here, the cross section of Fe K-shell ionization peaks at $\sim10$~MeV and $\sim20$~keV for protons and electrons, respectively \citep{Tatischeff12}.  The line intensity is simply in proportional to the density of LECRs and neutral Fe atoms.  
The Fe~\emissiontype{I} K$\alpha$ line that is most likely due to LECRs has been detected from more than 10 SNRs \citep{Sato14, Nobukawa18, Bamba18, Saji18a, Saji18b}. Most of them are known to be associated with molecular clouds (MCs; \cite{Jiang10, Sano15}), which means that they have a lot of targets for LECRs accelerated in the SNRs.  

IC\,443 is a middle-aged SNR (the age is thought to be $\tau_{\rm age}\sim3,000$--$30,000$~yr; \cite{Petre88, Olbert01}). MCs are interacting with the SNR \citep{Cornett77}. The GeV and VHE gamma-rays were detected close to the SNR-MC interaction region \citep{Albert07, Acciari09, Abdo10, Tavani10}.  Association between the gamma-rays and MCs and the spectral characteristic of pion decays   indicate the hadronic origin \citep{Abdo10, Ackermann13}. \citet{Indriolo10} measured H$_3^+$ column density in the vicinity of IC\,443 and found the high ionization rate of $\zeta_{2}\approx2\times10^{-15}$~s$^{-1}$, which is about five times larger than typical Galactic values. 

Those observational results suggest that IC\,443 accelerates a large amount of LECRs. In fact, \citet{Hirayama19}  analyzed the Suzaku high-statistics data of the northeast part of the SNR, and succeeded in discovering the Fe~\emissiontype{I} K$\alpha$ line from the spectrum. 
IC\,443 is known to exhibit the recombining plasma (RP; \cite{Yamaguchi09, Ohnishi14, Matsumura17, Hirayama19}).
In order to explain both the RP and the Fe~\emissiontype{I} K$\alpha$ line, the authors proposed  a scenario that the line is generated by LECRs.  Since IC\,443 is largely extended in appearance ($\sim\timeform{45'}$) due to its closeness ($\sim1.5$~kpc; \cite{Welsh03}) and  has been well studied in many wavelengths from radio to gamma-rays,  the SNR is a suitable target to develop  understanding of LECRs. In this paper, by utilizing all the Suzaku data of IC\,443, we  report further evidence for the Fe~\emissiontype{I} K$\alpha$ line induced by LECRs  and provide implications for acceleration of LECRs. 

\section{Observations and Data Reduction}
The observation logs are summarized in table~\ref{obs_log}. Suzaku observed three regions of IC443: the northeast, northwest, and south regions. \citet{Hirayama19} analyzed the northeast region. We utilized  all the data of IC\,443 obtained by  the X-ray Imaging Spectrometer (XIS;  \cite{Koyama07}). 
The XIS consists of four CCD cameras (XIS0, 1, 2, and 3). Each CCD is placed on the focal plane of the X-Ray Telescope (XRT; \cite{Serle07}). XIS0, 2, and 3 employ front-illuminated (FI) CCDs, and XIS1 has a back-illuminated (BI) CCD.  The field of view (FOV) of the CCD is $\timeform{17.8'} \times\timeform{17.8'}$.  The entire FOV of XIS2 and one-fourth of XIS0 have been out of function since 2006 November and 2009 June, respectively.
We reprocessed the data by using \texttt{xispi} in the analysis software package, HEAsoft 6.20, and the Suzaku calibration
database (CALDB) released in 2016 February, with the standard event selection criteria for the XIS data processing. The ancillary response  file (arf) and redistribution matrix file (rmf) were produced by \texttt{xissimarfgen} and \texttt{xisrmfgen} \citep{Ishisaki07}, respectively. The non-X-ray background (NXB) was estimated by \texttt{xisnxbgen} \citep{Tawa08}, and was subtracted from the X-ray images and spectra in this paper.

\begin{table*}
  \tbl{Observation logs.}{%
  \begin{tabular}{ccccc}
  \hline
 Obs.ID & \multicolumn{2}{c}{Pointing direction} 								& Observation start 	& Exposure \\
             & $\alpha_{\rm J2000.0}$ (\timeform{D}) 	& $\delta_{\rm J2000.0}$ (\timeform{D})  &  (UT) 	 	& (ks)  \\  
  \hline
501006010 & 94.2975 & 22.7757 & 2007-03-06 10:40:19 & 42.0 \\
501006020 & 94.2972 & 22.4797  & 2007-03-07 12:22:51 & 44.0 \\
505001010 & 93.9975 & 22.7552  & 2010-09-17 07:39:34 & 83.2 \\
507015010 & 94.2974 & 22.7535  & 2012-09-27 05:29:48  & 101.8 \\
507015020 & 94.3028 & 22.7465  & 2013-03-27 04:15:06  & 59.3 \\
507015030 & 94.3026 & 22.7461  & 2013-03-31 11:44:34  & 131.2 \\
507015040 & 94.3024 & 22.7479  & 2013-04-06 05:21:49  & 75.6 \\
  \hline
  \end{tabular}}\label{obs_log}
 \end{table*}

\section{Analysis}

\subsection{Fe~\emissiontype{I} K$\alpha$ intensity map}\label{sec:image}
\begin{figure*}
\centering
	\includegraphics[width=14cm]{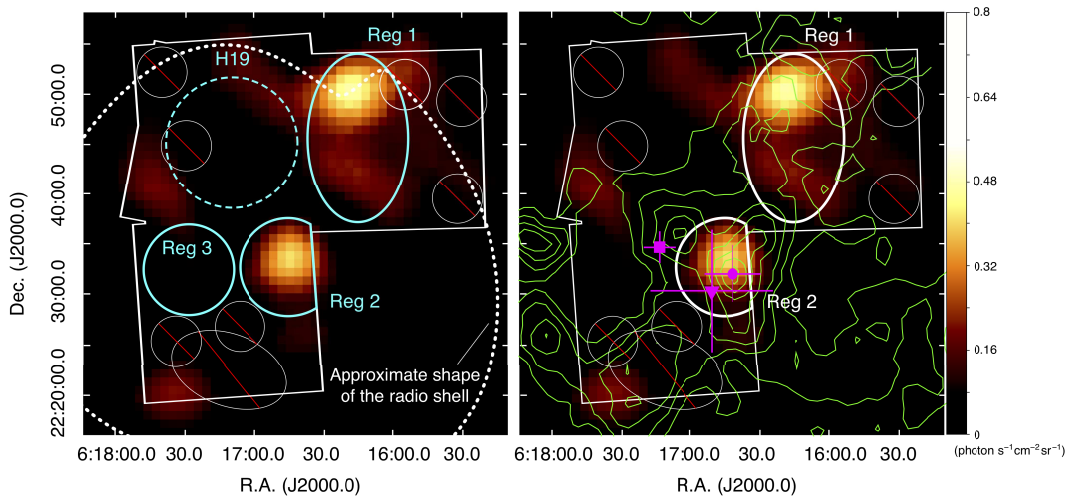}
 	 \caption{Left panel: intensity profile of Fe~\emissiontype{I} K$\alpha$ line. Color scale is in linear. Vignetting is corrected. The regions surrounded by white lines show the FOVs of the XIS. The dotted white line represents the approximate shape of the radio shell \citep{Lee08}. Point sources and the PWN 1SAX\,J0617.1$+$2221, which are marked with the solid white circles and the ellipse regions, respectively, are excluded from the image (see sec.\ref{sec:image}). The Fe~\emissiontype{I} K$\alpha$ line is enhanced in Regs~1 and 2 indicated by the solid cyan lines. We extracted X-ray spectra from the two regions, Regs~1 and 2, as well as Reg~3. \citet{Hirayama19} detected the Fe~\emissiontype{I} K$\alpha$ line in the H19 region shown by the dashed cyan  circle.
Right panel: Same as the left panel, but overlaid with the contours indicating the distribution of $^{12}$CO ($J=$1--0) at $-18$--$+10$~km~s$^{-1}$(Yoshiike 2019, private communication). The magenta marks and crosses represent the centroids of gamma-ray sources observed by Fermi (square), MAGIC (circle), and VERITAS (triangle) and their position errors \citep{Albert07, Acciari09, Abdo10}.  }\label{fig:image}
\end{figure*}

The left panel of figure~\ref{fig:image} shows the Fe~\emissiontype{I} K$\alpha$ line intensity map with the binning size of $48\times48$~pixels ($0.8\times0.8$~arcmin$^2$) and Gaussian smoothing  with $\sigma=3$~bins. Vignetting is corrected. The point sources and the pulsar wind nebula (PWN) 1SAX J0617.1$+$2221 reported in the previous works (\cite{Bocchino01}, \yearcite{Bocchino03}) are excluded from the image. The dotted line approximates the radio shell \citep{Lee08}. The notable structure of the intensity map is  bright blob-like enhancements located in the northwest (hereafter Reg~1) and the middle of IC\,443 (Reg~2). Especially, the north part of Reg~1 is the brightest enhancement, which is located near the radio shell.  The region where \citet{Hirayama19} detected the Fe~\emissiontype{I} K$\alpha$ line is indicated by the dashed cyan circle (hereafter H19).  
In the right panel of figure~\ref{fig:image}, we overlaid the distribution of $^{12}$CO ($J=$1--0)  (Yoshiike 2019, private communication) with the Fe~\emissiontype{I} K$\alpha$ line intensity map.  Reg~1 overlaps the elongate MC whereas Reg~2 coincides with the dense core of the MC.

\subsection{Spectral analysis}\label{sec:spec}
We extracted spectra from Regs~1 and 2,  as well as Reg~3  for reference (see figure~\ref{fig:image}). 
Since our objective is investigation of the Fe~\emissiontype{I} K$\alpha$ line, we  focused on the hard-band spectra (4--10~keV).  Here we utilized only the FI spectra because the BI data have poor signal-to-noise ratio above 7~keV due to the large NXB.  The fitting model consists of bremsstrahlung,  a power-law component with the Fe~\emissiontype{I} K$\alpha$ and K$\beta$ lines, the Fe~\emissiontype{XXV} and Fe~\emissiontype{XXVI} K$\alpha$ lines, and the cosmic X-ray background (CXB). 
Due to the broad point spread function of the XRT, the contamination of the flux from the PWN 1SAX J0617.1$+$2221 cannot be ignored, especially for  Regs~2 and 3 because of their proximity to the bright source. We simulated the contamination flux by {\tt xissimarfgen} by using the morphology and the flux of the PWN reported in \citet{Bocchino01}. 

 It was hard to  constrain the electron temperature ($kT_{\rm e}$) in the hard-band fitting. Previous works obtained $kT_{\rm e}\sim0.6$~keV  \citep{Yamaguchi09, Ohnishi14, Matsumura17, Hirayama19}, and thus we fixed $kT_{\rm e}$ to be 0.6~keV.  
Also we could not constrain photon index ($\Gamma$) of the power-law component, and $\Gamma$ is fixed to 2.5 in the same way as \citet{Hirayama19}.   The parameters of the CXB model are fixed to the values in \citet{Kushino02}. The line energies of the Gaussians were in principle free parameters for Fe~\emissiontype{I}, Fe~\emissiontype{XXV}, Fe~\emissiontype{XXVI} K$\alpha$ lines, but in the case that the constraint by fitting was difficult due to poor statistics, they were fixed to 6.40, 6.68, and 6.97 keV, respectively \citep{Kaastra93, Smith01}. The line center and intensity of the Fe~\emissiontype{I} K$\beta$ line are fixed to 7.06~keV and 0.125 times that of the Fe~\emissiontype{I} K$\alpha$ line \citep{Kaastra93}, respectively.  The line intensities of the other lines are free parameters. 

\begin{figure*}
    \begin{tabular}{cc}
      \begin{minipage}[t]{0.5\hsize}
        \centering
        \includegraphics[width=7cm]{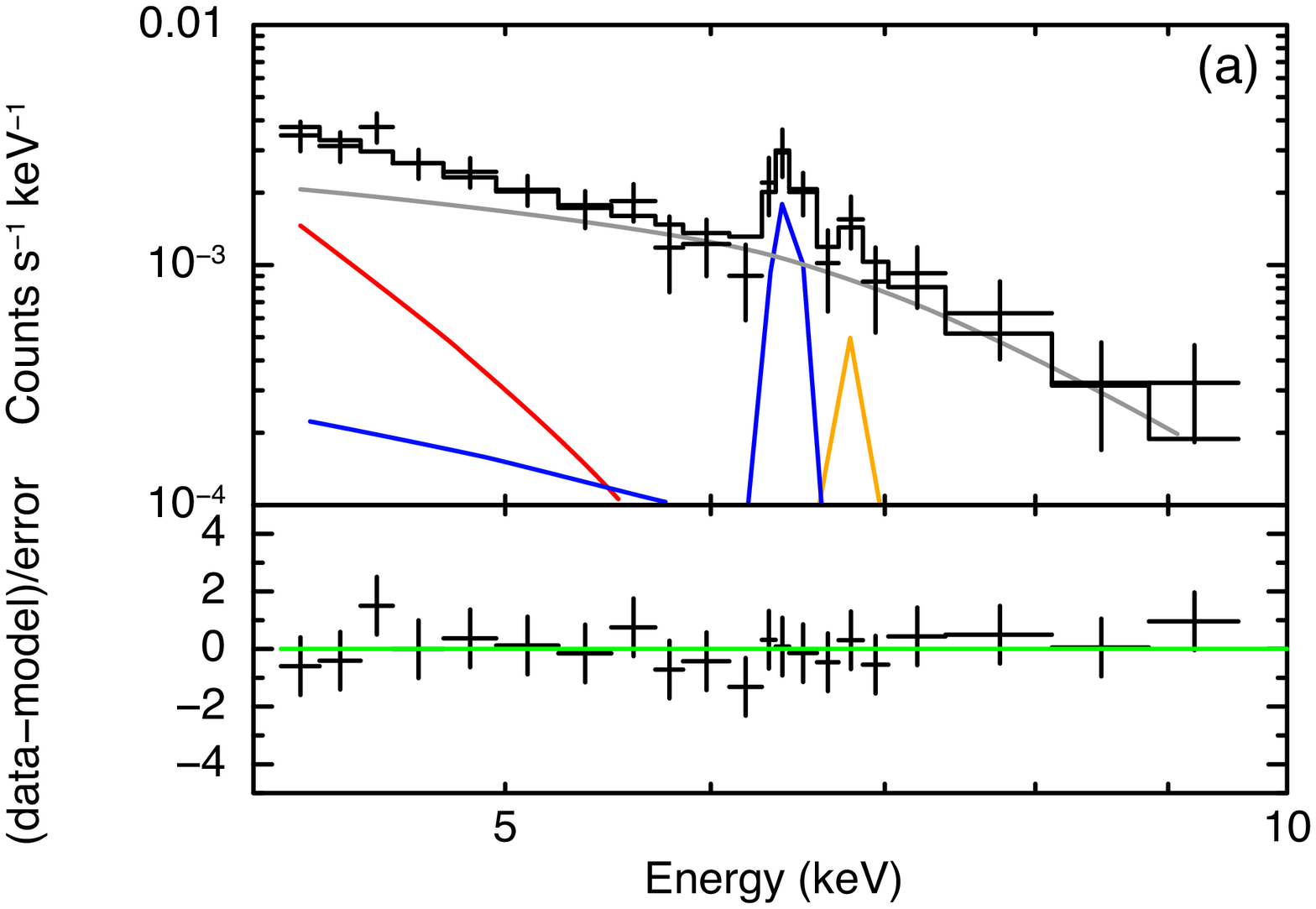}
      \end{minipage} &
      \begin{minipage}[t]{0.5\hsize}
        \centering
        \includegraphics[width=7cm]{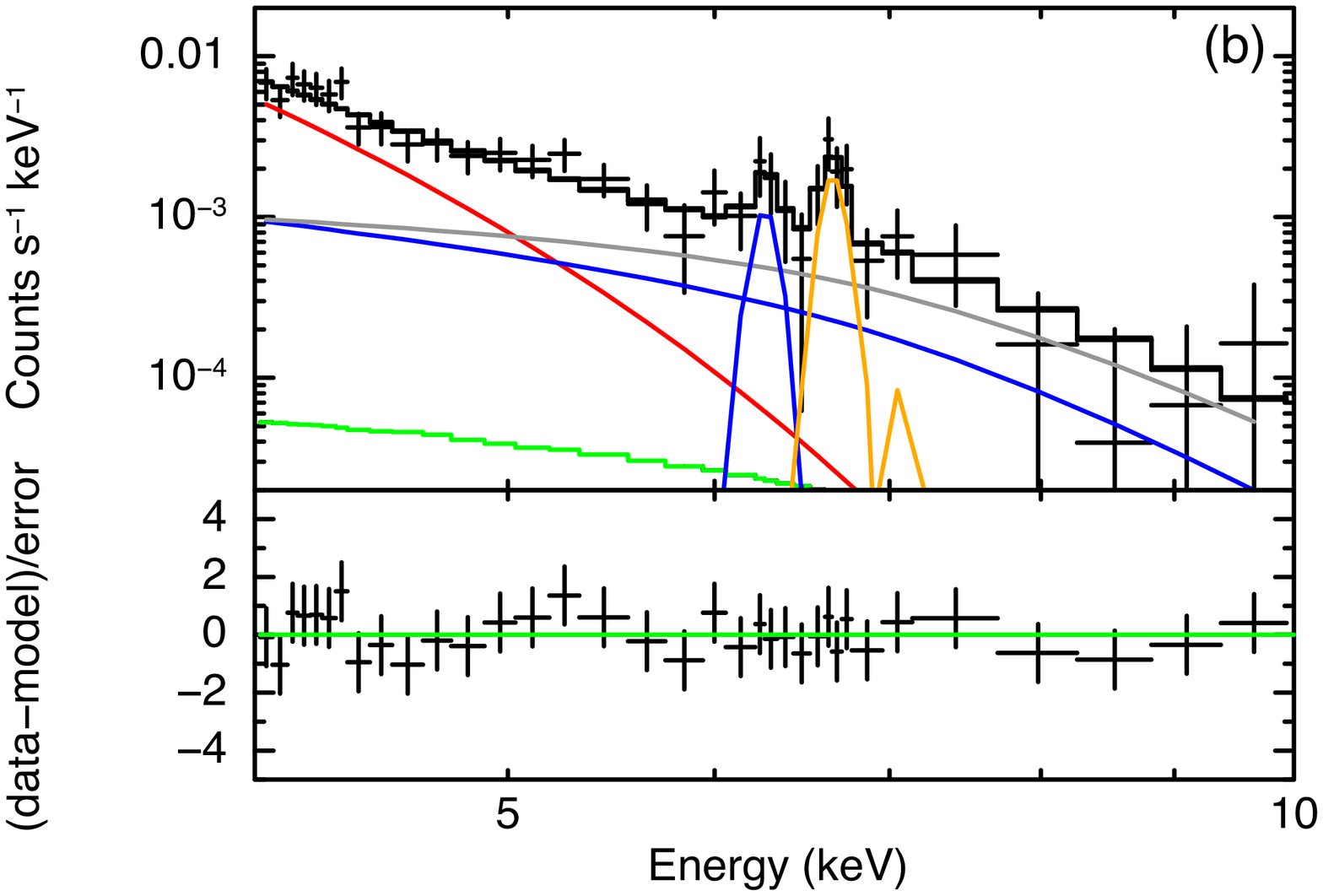}
      \end{minipage} \\
      \multicolumn{2}{c}{\begin{minipage}[t]{0.5\hsize}
        \centering
        \includegraphics[width=7cm]{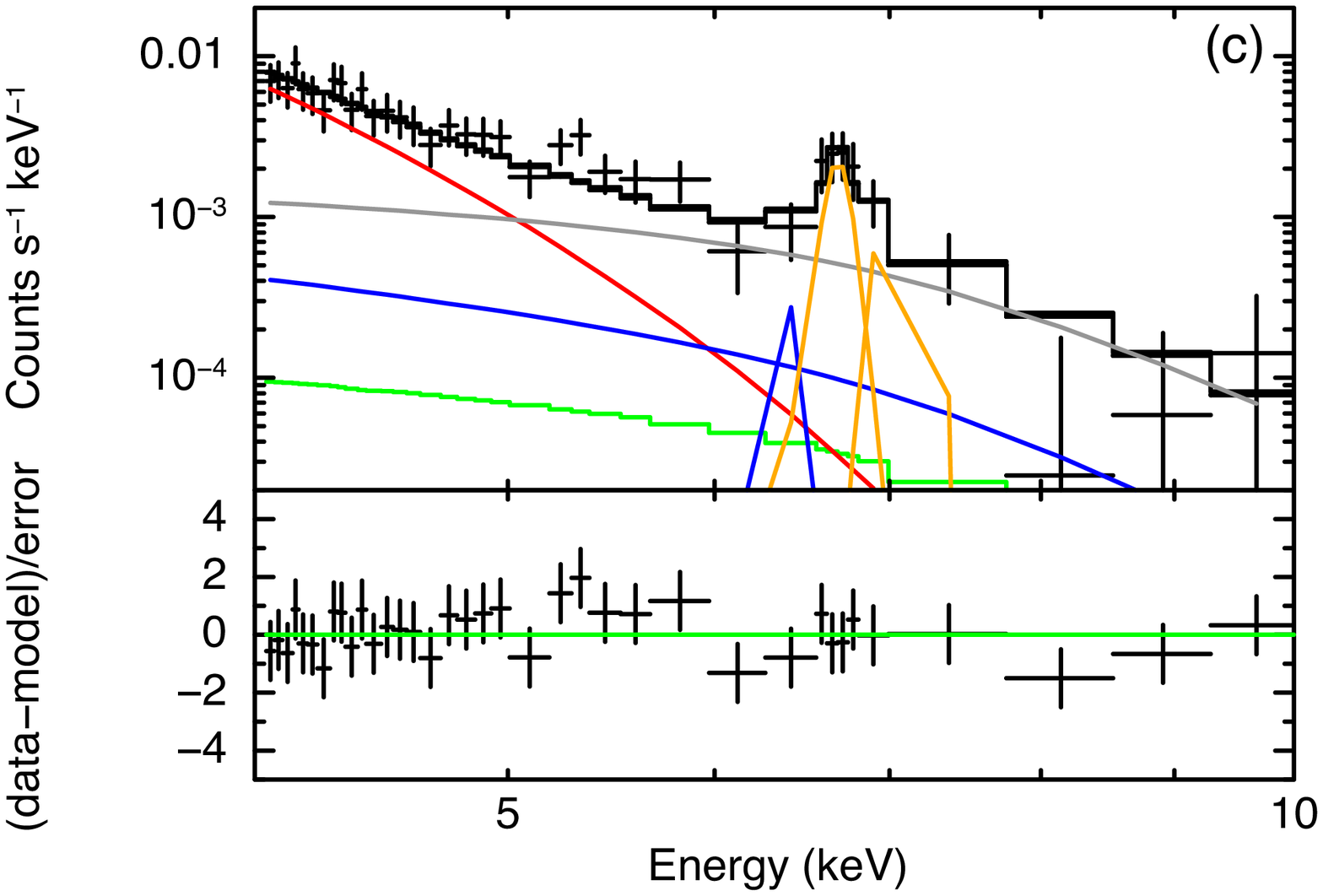}
      \end{minipage}}
    \end{tabular}
      \caption{X-ray spectra extracted from  Reg~1 (a), Reg~2 (b) and Reg~3 (c) and the best-fit models. The solid red, blue, and orange lines are bremsstrahlung, power-law component (power law plus the Fe~\emissiontype{I} K$\alpha$ line), Fe~\emissiontype{XXV} and Fe~\emissiontype{XXVI} K$\alpha$ lines, respectively. The gray lines indicates the CXB. The green lines in Reg~2 and Reg~3 shows the contamination flux from 1SAX J0617.1$+$2221. }\label{fig:spec}
  \end{figure*}

The intensity of the Fe~\emissiontype{I} K$\alpha$ line in the individual regions are  summarized in table~\ref{tab:intensity}. The value in H19 obtained by \citet{Hirayama19} is also displayed for comparison. 
The Regs~1 and 2 spectra show a clear Fe~\emissiontype{I} K$\alpha$ line as are shown in figure~\ref{fig:spec}, and their significance levels of  the line are 3.8$\sigma$ and 2.1$\sigma$, respectively. 
On the other hand, no significant Fe~\emissiontype{I} K$\alpha$ line is detected in  Reg~3, and we obtained the 90\% upper limit of the line intensity.  
For the Regs~1 and 2 spectra, we obtained the equivalent width (EW) of Fe~\emissiontype{I} K$\alpha$ line (the intensity ratio of the line to the power-law component): $>1.2$~keV and $0.7^{+0.9}_{-0.6}$~keV, respectively, with the 90\% confidence levels. As for Reg~3, the EW could not be constrained and thus is fixed to that in Reg~2. 
The above model explained all the spectra; the $\chi^2/$d.o.f. of Regs~1, 2, and 3 are $8/15= 0.53$, $16/29=0.55$, and $23/32=0.72$, respectively.

\begin{table*}
  \tbl{Centroids and intensity of the Fe~\emissiontype{I} K$\alpha$ line at 90\% confidence levels.}{%
  \begin{tabular}{lcc}
  \hline
 Region 		&  Centroids 	& Intensity 	\\
           		& (keV)		& (photons~s$^{-1}$~cm$^{-2}$~sr$^{-1}$) 				\\  
  \hline
Reg~1 		& $6.4\pm0.1$	& $0.273\pm0.119$	 \\
Reg~2		& $6.3\pm0.1$	& $0.344\pm0.277$		\\
Reg~3	 	& 6.4 (fixed)	& $< 0.345$	\\
H19\footnotemark[$*$] & $6.43\pm0.04$ & $0.079\pm0.035$		 \\
  \hline
  \end{tabular}}\label{tab:intensity}
 \begin{tabnote}
\footnotemark[$*$]  The values measured by \citet{Hirayama19}.
 \end{tabnote}
 \end{table*}

\section{Discussion}
\subsection{Origin of the Fe~\emissiontype{I} K$\alpha$ line}
We investigated the Fe~\emissiontype{I} K$\alpha$ emission in IC\,443 by utilizing the Suzaku data, and found the blob-like enhancement in  the northeast (Reg~1) and the middle (Reg~2) of the SNR. 
Regs~1 and 2 are associated with interaction sites between IC\,443 and MCs  (the right panel of figure~\ref{fig:image}). Reg~1 is close to a bright thermal X-ray knot which seems to mark the impact of a jet-like structure of ejecta with a dense cloud \citep{Greco18} and Reg~2 is associated with a sharp edge in the thermal X-ray emission, just at the position of the CO cloud (e.g. \cite{Troja06}).
We first discuss the origin of the Fe~\emissiontype{I} K$\alpha$ line.

One possible radiation process is the collisional ionization of Fe atoms in the ionizing plasma (IP), where the Fe-rich ejecta is in a low ionization state and the Fe K$\alpha$ line at $\sim6.4$~keV can be emitted (see \cite{Yamaguchi14}).  
In IC\,443, however, the Fe K$\alpha$ band is dominated by the RP  \citep{Yamaguchi09, Ohnishi14, Matsumura17, Hirayama19}. 
It is impossible that the RP of IC\,443 radiates a detectable  6.4~keV line. 
In addition, the morphologies of the line intensity map (figure~\ref{fig:image}) is totally different from that of the thermal plasma (e.g. \cite{Troja08}, \cite{Matsumura17}).
Those facts would also deny the plasma origin.   

Since the Fe~\emissiontype{I} K$\alpha$ line is associated with the MCs,   
another possible scenario is the inner shell ionization of neutral Fe atoms by X-rays from an external source or LECRs. 
The PWN 1SAX\,J0617.1$+$2221 can be a possible irradiating source.  
The required flux to explain the Fe~\emissiontype{I} K$\alpha$ line flux is $F_{\rm X}=5\times10^{-11} (4\pi/\Omega)(1\times10^{22}/N_{\rm H})$~erg~s$^{-1}$~cm$^{-2}$ in the 2--10~keV band. Here, $\Omega$ is a solid angle with which the MCs are seen from the PWN, and $N_{\rm H}$ is a hydrogen column density. The hydrogen column densities of the MCs associating with IC\,443 is ($0.6$--$1)\times10^{22}$~cm$^{-2}$  according to X-ray observations \citep{Troja06, Matsumura17, Hirayama19}. 
The solid angles of Regs~1 and 2 from the PWN are $\sim0.5$~sr and $\sim0.1$~sr, respectively. Therefore, the required flux is estimated to be $F_{\rm X}=(1$--$5)\times10^{-9}$~erg~s$^{-1}$~cm$^{-2}$.  
This is at least two orders of magnitude higher than the observed flux of 1SAX\,J0617.1$+$2221,  $F_{\rm X}=6\times10^{-12}$~erg~s$^{-1}$~cm$^{-2}$ \citep{Bocchino01}. 

Therefore, we propose that the Fe~\emissiontype{I} K$\alpha$ line  originates from LECRs that are produced by the SNR shock. 
Whether the LECRs are protons (with kinetic energy of the MeV band) or electrons (in the keV band) is distinguished by the EW of the Fe~\emissiontype{I} K$\alpha$ line.  
In electron bombardment, the EW ranges in the 0.2--0.4~keV while in the proton case the EW would be $>0.6$~keV \citep{Dogiel11}.  
We obtained the EW of $>1.2$~keV and $0.7^{+0.9}_{-0.6}$~keV for Regs~1 and 2, respectively (section 3.2). 
The scenario that can explain both the regions is MeV-proton bombardment.

\subsection{Comparison with Gamma-ray observations}\label{sec:Makino}
Since SNRs are expected to accelerate CRs with various energies, MeV CR 
protons that we detected through the Fe~\emissiontype{I} K$\alpha$ line should be
associated with higher-energy CRs that emit gamma-rays through proton-proton
interaction. For middle-aged SNRs, GeV and VHE gamma-rays are often
associated with  MCs. 
Since Reg~2  is close to the centroid of the gamma-ray clump 
(the right panel of figure~\ref{fig:image}), 
the MeV protons that generate the Fe~\emissiontype{I} K$\alpha$ line may be
connected to the HECRs. 

\begin{figure}
 \begin{center}
  \includegraphics[width=8.0cm]{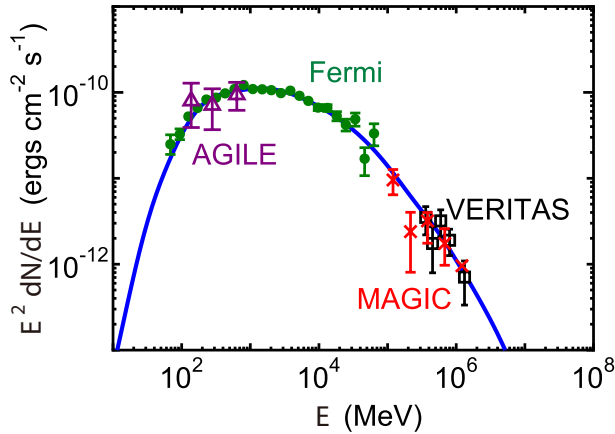}
 \end{center}
 \caption{Comparison of the best-fit model (solid line) with Fermi
(green filled circles; \cite{Ackermann13}), VERITAS (black open squares;
\cite{Acciari09}), MAGIC (red crosses; \cite{Albert07}), and AGILE
(purple open triangles; \cite{Tavani10}) observations for the SNR
IC~443.}\label{fig:gamma}
\end{figure}

Recently, \citet{Makino19} showed that both Fe~\emissiontype{I}
K$\alpha$ line and gamma-ray emissions from W28 and W44 can be explained
by a CR escaping model for SNRs. This model assumes that the SNRs are
interacting with MCs. LECRs are currently leaking from the SNRs into the
MCs and create Fe~\emissiontype{I} K$\alpha$ line emissions, while HECRs
had diffused out of the SNRs and generate gamma-rays through proton-proton
interaction. We applied this model to IC\,443; details of the model are
described in Appendix. In summary, the spectrum of CRs accelerated at
IC~443 is constrained by the gamma-ray spectrum
(figure~\ref{fig:gamma}). Using the results, we calculate the
Fe~\emissiontype{I} K$\alpha$ line intensity and find that it should be
$0.29^{+0.10}_{-0.11}$~photons~s$^{-1}$~cm$^{-2}$~sr$^{-2}$. The
predicted intensity is consistent with those for Regs~1 and 2 in
table~\ref{tab:intensity}. This supports the scenario that the observed
Fe~\emissiontype{I} K$\alpha$ line is associated with the LECR protons
that are accelerated along with the HECRs.

\section{Conclusion}
We discovered the bright blob-like enhancements of the Fe~\emissiontype{I} K$\alpha$ line located in the northwest and the middle of IC\,443. The most plausible origin of the Fe~\emissiontype{I} K$\alpha$ line is that the LECR protons accelerated in the SNR leak into the MCs and ionize the Fe atoms therein.   The observed line intensity is consistent with the value expected by the gamma-ray spectra and a CR escaping model.

\begin{ack}
We thank all the members of the Suzaku teams. 
We also thank Professor Katsuji Koyama for valuable comments. We are grateful to Dr. Hidetoshi Sano for providing the CO data shown in figure~\ref{fig:image}.  K.K.N. is supported by Research Fellowships of JSPS for Young Scientists. This work was supported by JSPS and MEXT KAKENHI Grant Numbers JP16J00548 (KKN), 18K03647 (YF) and JP17K14289 (MN).
\end{ack}

\appendix
\section*{Application of the CR escaping model to Gamma-rays and Fe~\emissiontype{I} K$\alpha$ line in IC\,443 }
We assume that MCs with density $n_{\rm H}$ are distributed
around the SNR in a shell region between $r=L_1$ and $r=L_2$ with a
filling factor $f_{\rm gas}$ (see figure~1b of \cite{Makino19}). From CO
observations with NANTEN2, we assume that $n_{\rm H}=730\:\rm cm^{-3}$,
$L_1=12$~pc, $L_2=15$~pc, and $f_{\rm gas}=0.05$ (Yoshiike 2019, private
communications). For these parameters, the mass of MCs is
$6.2\times 10^3\: M_\odot$.  The distance to IC~443 is assumed to be
1.5~kpc \citep{Welsh03}. The SNR contacts with the MCs at the
radius $L_1$. The SNR is assumed to be in the Sedov phase and the
evolution is the same as those for W28 and W44 \citep{Makino19}. Thus,
the current time is $t_{\rm obs}=1.5\times 10^4$~yr after the supernova
explosion. The momentum of escaping CRs, $p_{\rm esc}$, is a decreasing
function of the shock radius $R_{\rm sh}$:
\begin{equation}
\label{eq:pesc}
 p_{\rm esc}(t) = p_{\rm max}
\left(\frac{R_{\rm sh}(t)}{R_{\rm Sedov}}\right)^{-\alpha}\:,
\end{equation}
where $p_{\rm max}$ and $R_{\rm Sedov}$ are the escape momentum and the
shock radius at the beginning of the Sedov phase ($t=t_{\rm Sedov}$),
respectively, and $\alpha$ is the index that is empirically derived. We
assume that $R_{\rm Sedov}=2.1$~pc, $t_{\rm Sedov}=210$~yr and
$\alpha=6.5$ as we did in \citet{Makino19}. At present ($t=t_{\rm
obs}$), CRs with momenta of $p>p_{\rm esc}(t_{\rm obs})$ have gradually
escaped from the SNR, while those with $p<p_{\rm esc}(t_{\rm obs})$ have
been confined around the shock front. However, since the shock collides
with MCs at $t\sim t_{\rm obs}$ and the MCs prevent the CR confinement,
the lower-momentum CRs are now escaping altogether. As a result, the CR
momentum spectrum have a break at $p=p_{\rm esc}(t_{\rm obs})$, which
results in a bend in the gamma-ray spectrum at $E\sim$~GeV (see
figure~\ref{fig:gamma}).

For the above parameters for the MCs and the SNR, we calculate gamma-ray
spectra and compare them with the observed ones. Then, we constrain our
model parameters for CRs through $\chi^2$ fitting.   In
\citet{Makino19}, the fitting parameters are the total CR energy $E_{\rm
CR,tot}$, the index $s$ of the CR momentum spectrum at the shock front
($\propto p^{-s}$), and the maximum momentum $p_{\rm max}$
[equation~(\ref{eq:pesc})]. In this study, we also treat parameters
associated with CR diffusion as fitting parameters because of the
following reasons.

The diffusion of CRs outside the SNR is calculated based on the model of
\citet{Ohira11}. We assume that the diffusion coefficient has a form
given by 
\begin{equation}
\label{eq:1}
D(p)=10^{28} \chi \left(\frac{pc}{10~\rm{GeV}}\right)^{\delta}\rm\: cm^{-2}\: s^{-1}, 
\end{equation}
where $\chi$ ($<1$) is a constant that accounts for a possible
suppression of the diffusion coefficient around SNRs (\cite{Fujita09},
\yearcite{Fujita10}, \yearcite{Fujita11}).
\citet{Makino19} adopted $\chi=0.5$ and $\delta=1/3$ for W28 and
W44. However, we found that these parameters cannot reproduce the convex
shape of the gamma-ray spectrum of IC~443 at $\gtrsim 1$~GeV
(figure~\ref{fig:gamma}). We note that the gamma-ray spectra for W28 and
W44 are approximately represented by a power-law at $\gtrsim 1$~GeV
\citep{Makino19}. The spectrum shape for IC~443 may indicate that the
diffusion coefficient or $\chi$ is much smaller than that for W28 and
W44. If $\chi$ is much smaller than 0.5, lower-energy CRs remain close
to the shock front and their distribution depends on the finite size of
the SNR. On the other hand, higher-energy CRs can diffuse away
relatively fast because the diffusion coefficient is an increasing
function of energy for $\delta>0$ [equation~(\ref{eq:1})]. Thus, their
distribution is not much different from the one when the SNR is
approximated as a point CR source. The boundary of energy creates
another break in the CR momentum spectrum, which is referred to as
$p_{\rm br,ext}$ in \citet{Ohira11}.  In this study, we treat
$\chi$ and $\delta$ as fitting parameters.

Through the parameter fitting, we obtain $E_{\rm
CR,tot}=3.1^{+0.5}_{-0.5}\times 10^{50}$~erg, $s=2.52^{+0.10}_{-0.20}$,
$p_{\rm max}c=4.0^{+3.9}_{-2.0}\times 10^{13}$~eV, $\chi=0.011^{+0.003}_{-0.002}$ and $\delta=0.58^{+0.07}_{-0.08}$, and the best-fit
result is shown in figure~\ref{fig:gamma}. From the results, we also
obtain $p_{\rm esc}(t_{\rm obs})=6.2^{+6.2}_{-3.1}\times 10^8$~eV, which
corresponds to the peak of the gamma-ray spectrum in
figure~\ref{fig:gamma}. The small diffusion coefficient we adopted
results in another break in the gamma-ray spectrum at $E\sim 10$~GeV
because $p_{\rm br,ext}c\sim 50$~GeV (figure~\ref{fig:gamma}).
Moreover, we find that the intensity of the
Fe~\emissiontype{I} K$\alpha$ line is $I_{\rm
6.4keV}=0.29^{+0.10}_{-0.11}\rm\: photons~s^{-1}\: cm^{-2}\: sr^{-1}$.
The line intensity is consistent with our observational results
(Regs~1 and 2 in table~\ref{tab:intensity}). 

The gamma-ray maxima appear to coincide with Reg 2
(figure~\ref{fig:image}). The distribution function of escaped HECRs is
generally represented by a gaussian, $\propto \exp(-r^2/r_{\rm
diff}^2)$, where $r$ is the distance from the SNR center and $r_{\rm
diff}$ is the diffusion radius (e.g. equation~(9) of
\cite{1995PhRvD..52.3265A}). Since $R_{\rm sh} < r_{\rm diff}$, the HECR
density is almost constant for $R_{\rm sh} < r < r_{\rm diff}$. In
our model, the gamma-rays are emitted from the MCs in that region and
Fe~\emissiontype{I} K$\alpha$ line is observed in MCs that are in touch
with the SNR ($r\sim R_{\rm sh}$). Thus, it is likely that gamma-rays
and the Fe~\emissiontype{I} K$\alpha$ line are observed in the same
MCs. Moreover, projection effects may also be responsible for the
spatial co-location of the gamma-rays and the Fe~\emissiontype{I}
K$\alpha$ line.

Assuming a point CR source, we can discuss the diffusion
coefficient based on the size of the Fe~\emissiontype{I} K$\alpha$ line
emission region for Reg~2.  The diffusion time $t_{\rm diff}$ of protons
is given by $t_{\rm diff} \sim R^2/6D'$, where $D'$ is the diffusion
coefficient and $R$ is the size of the Fe~\emissiontype{I} K$\alpha$
line emission.  Figure~\ref{fig:image} indicates that the angular sizes
of Reg~2 is $\timeform{4'}$, which corresponds to $R=2$~pc (assuming
that the distance of IC\,443 is 1.5~kpc; \cite{Welsh03}).  According to
the equations (4.24) of \citet{Mannheim94} and the equation (C3) of
\citet{Strong98}, the ionization cooling time of 10~MeV protons is
estimated to be $t_{\rm cool}\sim500$~yr for Reg~2 with
$n_H=730$~cm$^{-3}$ (Yoshiike 2019, private communication). The
diffusion coefficient is required to be $D'\geq 4\times10^{26}$
~cm$^{-2}$~s$^{-1}$ from the condition of $t_{\rm diff}\leq t_{\rm
cool}$.  This is much larger than the coefficient at the kinetic energy
of 10~MeV ($D(p\approx 137{\rm MeV/c})=9\times10^{24}$
~cm$^{-2}$~s$^{-1}$) given by equation~(\ref{eq:1}) with $\chi=0.011$
and $\delta=0.58$.

We note that even if the reality is $D'\gg D(p)$, it would be
allowed.  In that case, the apparent inconsistency may show the
difference of the diffusion coefficient between the inside and the
outside of the MC. Moreover, the CRs may not be injected into the MC
from a point source.  For example, the size of the Fe~\emissiontype{I}
K$\alpha$ line emission may reflect the area of interaction between the
shock and the MC, from which the CRs are injected.

\end{document}